# The Five Elements of Th1-Th2 System

Shengrong   Zou*

*Information Engineering College, Yangzhou University , Yangzhou 225009 JiangSu ,PR China*

___________________________________________________________________________

**Abstract**

The T helper (Th) phenotypes, Th1/Th2, are acquired upon interaction of a naive T helper cell and an antigen presenting cell (APC). Naive T helper cells may differentiate into either phenotype, and the actual outcome is determined by the density and avidity of the antigenic determinants presented by the APC, and the APCs inherent costimulatory properties. Until recently it was thought that differentiation is further affected by cytokines. In our work, We have specified aspects of   T-cell cytokine networks   using B method of software engineering. With this model, we are able to run verification with B-toolkit and allow us to compare the dynamic behavior of the model to actual experimental data from College of Animal Science and Veterinary Medicine. Here we present a Wu-Hsing model of Th1-Th2 system.

___________________________________________________________________________

**Contents**

___________________________________________________________________________

* Tel: +86 514 87339226 ; fax : +86 514 7887937.
  *E-mail address:* Zoushengrong@vip.sina.com(SR. Zou).

# 1 Introduction

Although many details of particular cytokine interactions have been elucidated and the effects of cytokines on a myriad of cellular functions have been described, practically nothing is known about the behavior of the network as a whole[1]. Perhaps the most important features of the network are nonlinearities in cytokine interactions and the presence of positive and negative feedback . Complex nonlinear systems commonly have unusual and nonintuitive properties that may include chaotic behavior. These properties make the cytokine network too complex to be understood fully by the conventional experimental approach of testing the effects of cytokines or combinations of cytokines on cells in vitro.

The finding that T cell immune responses could be divided into those promoting cell mediated immunity (Th1) and humoral responses (Th2) has had a profound effect on the understanding of immune response generation. With ever increasing knowledge of the immune system, the model has come under criticism, as not all responses easily fit the classification. All cytokine interactions exhibit nonlinear behavior. In fact , they act in a complex, intermingled network where one cytokine can influence the production of, and response to, many other cytokines[2]. A complementary modeling approach based on modern nonlinear dynamic is now required. In this review we update the model with current thinking regarding the generation and maintenance of immune responses[3].

# 2 The Wu-Hsing model

The thinkers of the early Zhou dynasty attempted to fuse many of the strains of Chinese thought to come up with a syncretic and systematic explanation of the universe, the changes that occur in the universe, and the relation of the human world to the physical and divine worlds. Their thought focused on two inventions, both designed to explain the changing world. The principles of yin and yang, opposite forces of change which complement and cyclically give rise to one another, operated through the physical mechanism of "the five material agents," or Wu-Hsing .

These five material agents are wood-fire-earth-metal-water and are grouped either in the order by which they produce one another (wood gives rise to fire, fire gives rise to earth, earth gives rise to metal, metal gives rise to water, water gives rise to earth, etc.) or the order by which they are conquered by one another: fire is conquered by water, water is conquered by earth, earth is conquered by wood, wood is conquered by metal, and metal is conquered by fire, etc. Each of these orders can be used to explain the progression of change in just about everything. The five agents, however, is a metaphysical explanation of the progression of change that is meant to be applied to every phenomenon one encounters in this changing universe: politics, ethics, music, biology, time, seasons, history, etc.

The Chinese divided the year into five seasons: Spring, Summer, Late Summer, Autumn, and Winter. Each of them they described with the attributes of one of the Five Elements: Wood, Fire, Earth, Metal, and Water. When looking at these Five Elements -- or Five Forces -- all around us in Nature, we can also discern that these qualities of energy have a matching counterpart inside every living being. These forces are the elemental building blocks of our inner nature.

In human anatomy, the spleen is ruled by wood, the lungs by fire, the heart by earth, the liver by metal, and the kidneys by water. If one has a disease of the liver, it is because the liver is being overcome by a fire agent or pathogen---since fire is overcome by water, one would treat the liver pathogen with a water agent.

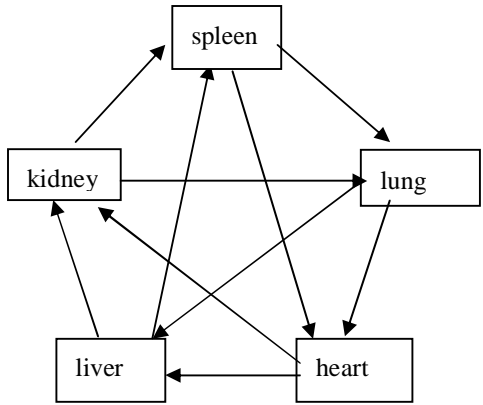

Fig.1. Five elements in Chinese medicine

One could endlessly list how the various categories of phenomenon fit into Wu-Hsing model. What is important to understand is that the five agents explain everything including the progress of change in the universe.

## 3 The Five Elements of Th1-Th2 system

A great deal of experimental data on the regulation of Th1 and Th2 differentiation have been obtained. Other authors[4-8] have modeled the Th1-Th2 system, with a variety of approaches and areas of emphasis[9,10]. But many essential features of this complex system are still not understood[11]. Here we present a Wu-Hsing model of Th1/Th2 differentiation and cross regulation.

Th1 and Th2 cells arise from a common precursor. When an immune response is initiated, naive helper T cells produce interleukin (IL)-2 and proliferate. After entering the cell cycle, progeny become competent to produce effector cytokines[12,13]. Th1 cells, but not Th2 cells, secrete IL2, interferon(IFN)-γ, and tumor necrosis factor(TNF)-β, whereas Th2 cells, but not Th1 cells, express IL4, IL5, IL6, and IL10.

The relative balance of Th1 vs. Th2 cytokine expression is thought to play a critical role in the regulation of cellular immune responses, with impacts on susceptibility to infectious disease and/or progression of inflammatory disorders [14].

The fact that some of these cytokines have been designated type-1 or type-2 factors does not imply that these cytokines cannot be produced also by other cells. Producer cells other than T-cells expressing CD4 include CD8 (+)T-cells, monocytes, natural killer cells, B-cells, eosinophils, mast cells, basophils, and other cells. This is why many immunologists define immune responses by the types of cytokines controlling these responses rather than by the types of cells.

Both types of T-helper cells can influence and regulate each other by the cytokines they secrete [15]. For example, IFN-γ, secreted by Th1 cells, can inhibit the proliferation of Th2 cells. IL10, secreted by Th2 cells, can suppress Th1 functions by inhibiting cytokine production. The Th2 cytokine IL4 inhibits the differentiation and/or expansion of Th1 cells. It thus appears that these functional subsets of helper cells are mutually antagonistic such that the decision of which subset predominates within an infection may determine also its outcome[16]. Through the activities of the cytokines produced, Th1 and Th2 cells can keep each other in check and prevent inflammatory reactions in response to pathogens getting out of control.

The main interactions influencing the generation and maintenance of Th1-Th2 system are summarized in Fig.2.

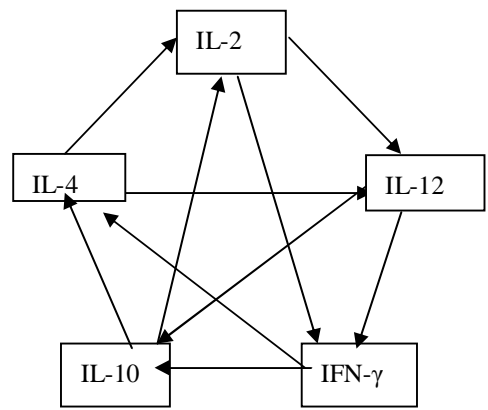

Fig.1. Cytokines are major inducers of Th1 and Th2 subset development. The five elements of Th1-Th2 system are IL-2, IL-12, IFN-γ, IL-4 and IL-10. IL-2 promotes production of IL-12, IL-12 promotes production of IFN-

ɣ, IFN-ɣ promotes production of IL-10, IL-10 promotes production of IL-4, IL-4 promotes production of IL-2. IL-10 inhibits production of IL-2, IL-2 inhibits production of IFN-ɣ, IFN-ɣ inhibits production of IL-4, IL-4 inhibits production of IL-12, IL-12 inhibits production of IL-10.

*3.1 Five elements*

IL-2 is a cytokine with a crucial regulatory role in the immune system. In healthy humans (or mammals), expression of IL-2 is restricted to the effector subset of T-helper lymphocytes—that is, activated naive CD4 T-helper cells and Th1-committed T-helper clones. IL-2 acts as a T-cell growth factor; it sensitizes activated T cells to activation-induced cell death (AICD); it promotes the maturation of cytotoxic T lymphocytes and lymphokine-activated killer activity; and it enhances germicidal and cytotoxic function of natural killer (NK) cells, B-cells, monocytes, and macrophages.

IL-12 is a regulatory protein produced by activated B lymphocytes and macrophages. The biological activities of IL-12 include the stimulation of growth of activated CD4 and CD8+T cells and NK cells. Human IL-12 also promotes the development of proinflammatory/Th1-like CD4+ cells and cytotoxic CD8 T cells.

IFN-ɣ is an acid-labile interferon produced by CD4 and CD8 T lymphocytes as well as activated NK cells. IFN-ɣ receptors are present in most immune cells, which respond to IFN-ɣ signaling by increasing the surface expression of class I major histocompatibility complex(MHC) proteins. This promotes the presentation of antigen to T-helper (CD4+) cells. IFN-ɣ signaling in APC and antigen-recognizing B and T lymphocytes regulate the antigen-specific phases of the immune response. Additionally, IFN-ɣ stimulates a number of lymphoid cell functions including the anti-microbial and anti-tumor responses of macrophages, NK cells, and neutrophils.

IL-4 is a pleiotropic cytokine that regulates diverse T and B cell responses including cell proliferation, survival and gene expression. Produced by mast cells, T cells and bone marrow stromal cells, IL-4 regulates the differentiation of naive CD4+ T cells into helper Th2 cells, which favor a humoral immune response. Another dominant function of IL-4 is the regulation of immunoglobulin class switching to the IgG1 and IgE isotypes. Excessive IL-4 production by Th2 cells has been associated with elevated IgE production and allergy.

IL-10 is an immunosuppressive cytokine produced by a variety of mammalian cell types including macrophages, monocytes, T cells, B cells and keratinocytes. IL-10 inhibits the expression of proinflammatory cytokines such as IL-1 and TNF-α. Like IL-4, IL-10 enhances humoral immune responses and attenuates cell-mediated immune reactions.

*3.2 Promotion*

IL-2---IL-12  IL-2 promotes the proliferation, differentiation, and survival of target cells. Subsequent to their expansion, the differentiated function of both T helper cells and T cytolytic cells is dependent upon an adequate supply of IL-2. [17,18].

IL-12---IFN-ɣ  IL-12 is especially important because its expression during infection regulates innate responses and determines the type and duration of adaptive immune response. IL-12 induces IFN-γ production by NK, T cells, dendritic cells (DC), and macrophages. IL-12 promotes the differentiation of naive $CD4^+$ T cells into Th1 cells that produce IFN-ɣ and aid in cell-mediated immunity[19].

IFN-ɣ---IL10 Recombinant murine IFN-ɣ was found to possess B cell maturation factor activity for resting splenic B cells and the comparable B cell tumor line WEHI-279.1. IFN-ɣ may be one of several molecules with a direct role in driving the maturation of resting B cells to active immunoglobulin secretion[20].

IL-10---IL-4  IL-10 inhibits production of IL-2 and TNF-α, but not IL-4 when T cells were stimulated without APCs [21]. IL-10 acts as a costimulator of the proliferation of mast cells (in the presence of IL-3 and/or IL-4 ) and peripheral lymphocytes.

IL-4---IL-2  IL-2 and IL-4 are important growth and differentiation factors for B and T cells. IL-4 antagonizes the effect of IL-2 on B cells and some T cells while it synergizes with IL-2 on other T cells[22].

*3.3 Inhibition*

IL-10---L-2  IL10 , secreted by Th2 cells, can suppress Th1 functions by inhibiting cytokine production. IL10 inhibits the synthesis of a number of cytokines such as IFN-ɤ, IL2 and TNF-β in Th1 T-helper subpopulations of T-cells[23].

IL-2---IFN-ɤ  IL-2 can promote expression of CTLA-4, a negative regulator of TCR signals, which competes with the positive, costimulatory molecule, CD28. In addition to positive influences attributable to IL-2, there are negative feedback regulatory effects of IL-2 that function to limit the ultimate immune response and inhibit production of IFN-ɤ [24].

IFN-ɤ ---IL-4  IFN-ɤ directly suppresses IL-4 gene expression. IRF-1 and IRF-2 induced by IFN-γ bind to three distinct IL-4 promoter sites and function as transcriptional repressors. Data demonstrate a direct negative feedback of IFN-ɤ on expression of the Th2 cytokine gene IL-4 [25].

IL-4---IL-12  Development of Th1 responses can be antagonized directly by IL-4 and indirectly by IL-10. IL-4 inhibits the production of inflammatory mediators such as IL-12 and IL-18 from macrophages and/or dendritic cells stimulated by the innate immune response[26,27].

IL-12 ---IL10  IL-12 promotes Th1 responses by inducing IFN-ɤ from T cells and NK cells. IL-12 inhibits the development of IL-4-producing Th2 cells and the production of IL-10[28].

# 4  Conclusion

The crucial cell for immune system control is the T-cell. In Th1-Th2 system, two types of T-helper cells have been defined on the basis of their cytokine secretion patterns[29]. The decision of a naive T cell to differentiate into Th1 or Th2 is crucial, since to a first approximation it determines whether a cell-mediated or humoral immune response is triggered against a particular pathogen, which profoundly influences disease outcome[30]. In our work, We have specified aspects of T-cell cytokine networks using B method of software engineering. With this model, we are able to run verification with B-toolkit and allow us to compare the dynamic behavior of the model to actual experimental data from College of Animal Science and Veterinary Medicine. This study shows that IL-2, IL-12, IFN-ɤ, IL-10, IL-4 are crucial regulator of Th1-Th2 system. This model therefore may have crucial significance in the development of therapeutic strategies for bio-pharmacologic intervention in cytokine-mediated inflammatory process and infections.

**Acknowledgements**

The research is supported by National Natural Science Foundation of China under the grant No. 70671089 and 10635040.